\documentclass[12pt,draftclsnofoot,journal,onecolumn]{IEEEtran}
\IEEEoverridecommandlockouts
\usepackage{algorithm}
\usepackage{amsmath}
\usepackage{algpseudocode}
\usepackage{graphicx}
\usepackage{wasysym}

\usepackage{mathtools}
\usepackage{enumitem} 
\usepackage{ifpdf}
\usepackage{amsthm}
\usepackage{amsmath}
\usepackage{nicefrac}
\usepackage{cite}
\usepackage{amsfonts}
\usepackage{comment}
 \usepackage{url}
 \usepackage{amssymb}
 \usepackage[paper=letterpaper,margin=0.75in]{geometry}
 \usepackage[ansinew]{inputenc}

\usepackage{lipsum}
\usepackage{hyperref}

\theoremstyle{definition}



\begin{document}
\title{Efficient and Privacy-preserving Voice-based Search over mHealth Data}

\author{\IEEEauthorblockN{Mohammad Hadian$^1$, Thamer Altuwaiyan$^1$, Xiaohui Liang$^1$, and Wei Li$^2$}\\
	\IEEEauthorblockA{$^1$Department of Computer Science, University of Massachusetts Boston\\
	}
	\IEEEauthorblockA{$^2$School of Computer and Information Engineering, Xiamen University of Technology\\
	}
	Email: \{mshadian,thamerfa,xiaohui\}@cs.umb.edu; liweipla@sina.com

\thanks{\copyright Personal use of this material is permitted. Permission from IEEE must be obtained for all other uses, in any current or future media, including reprinting/republishing this material for advertising or promotional purposes, creating new collective works, for resale or redistribution to servers or lists, or reuse of any copyrighted component of this work in other works.}

}

\date{}
\maketitle
\thispagestyle{plain}
\pagestyle{plain}

\newtheorem{definition}{Definition}

\begin{abstract}
	In-home IoT devices play a major role in healthcare systems as smart personal assistants. They usually come with a voice-enabled feature to add an extra level of usability and convenience to elderly, disabled people, and patients. In this paper, we propose an efficient and privacy-preserving voice-based search scheme to enhance the efficiency and the privacy of in-home healthcare applications. We consider an application scenario where patients use the devices to record and upload their voice to servers and the caregivers search the interested voices of their patient's based on the voice content, mood, tone and background sound. Our scheme preserves the richness and privacy of voice data and enables accurate and efficient voice-based search, while in current systems that use speech recognition the richness and privacy of voice data are compromised. Specifically, our scheme achieves the privacy by employing a homomorphic encryption; only encrypted voice data is uploaded to the server who is unable to access the original voice data. In addition, our scheme enables the server to selectively and accurately respond to caregiver's queries on the voice data based on voice's feature similarity. We evaluate our scheme through real experiments and show that our scheme even with privacy preservation can successfully match similar voice data at an average accuracy of 80.8\%.
\end{abstract}


%
\IEEEpeerreviewmaketitle

\section{Introduction}
Recently IoT in-home devices have shown their potential to significantly improve in-home healthcare system by further facilitating the patient-caregiver relationships~\cite{holopainen2007use}. For instance, hands free and device free voice command features have significantly helped elders, disabled people and patients. Voice reminders and notifications for taking medications or doctor appointments, and ease of communication with caregivers through voice enabled hands-free messaging services, are some examples of applications helping everyone in the healthcare cycles~\cite{ceer2006pervasive}. In current applications of voice-enabled IoT devices, the voice data are usually converted to text mainly because i) the required capacities for storing text is significantly less than voice and ii) data mining and extraction from text is much easier and efficient than voice~\cite{schober2015precision}. However, there are three main problems with speech recognition in health related applications. \textbf{First,} voice is a richer date type than text, i.e. there exist useful information that are specific to voice, like tone, volume, pitch~\cite{forsberg2003speech} and this richness cannot be preserved by speech recognition. As shown in~\cite{online}, richness of the voice is important for healthcare applications. \textbf{Second,} considering the healthcare data, as a clearly sensitive and private category of data, and the home environment, which is always known as the most private environment for users, there has always been a challenge for keeping the data private and secure from other parties outside the trust cycle, e.g. untrusted servers, and outsourcing services. The speech recognition requires high processing and storage resources because of its complexity. It is thus usually done on cloud servers, while this requirement limits the application in case of sensitive information and untrusted servers. \textbf{Finally,} the offline speech recognition techniques are a solution to the privacy preservation problem on untrusted servers, but they are less accurate compared to the online solutions. Also, the training phase in speech recognition may limit the application in case of dynamic and continuously growing systems~\cite{mcgraw2016personalized}.

In this paper we consider a scenario where voice data are generated by the user, and the information contained in these voice data such as tone, background voices and ambient sounds are being detected and utilized. Richness of the voice is important for health related applications. The background voices can reveal information about the environment, e.g., a music being played, a show on TV, or presence of other people. 
Moreover, the tone of a patient's voice can easily and clearly reveal some information about her emotional and physical conditions and reflect her feelings and mood. In addition, happiness, sadness, anger or frustration can also be detected in patient's voice even if the patient says the same word. To preserve the richness and privacy of the voice data, we propose an efficient and privacy-preserving voice-based search scheme, which stores the patient's voices collected through voice-enabled IoT home devices at a server and enables the caregiver to later search their interested voices from the server. To preserve data privacy at the server, encrypted data storage is one popular technique ~\cite{boneh2004public}. This technique is highly implemented and used for text, images, and video, but there is a small amount of prior work in case of voice data because of popularly used speech-to-text conversion services nowadays. The original voices are not usually stored in the database, and are mostly kept in special cases where the voice itself is important, e.g. music sound records. In our scenario, these voices will be encrypted and uploaded to a server and the patient's caregiver can later query the interested data from the server, decrypt it and access the original voice of the patient. We also aim to achieve higher accuracy than existing works because it deals with the voice data directly and does not need to convert them to text. Specifically, contributions of this work can be summarized as follows:
\begin{itemize}
	\item First, we study the advantages and disadvantages of voice over text as the data type to be collected from the patients, from both patient and caregiver's point of view. We found the richness of voice can be useful to in-home healthcare applications where existing data collection and search schemes cannot be applied.
	\item Second, we present novel schemes for collecting, encrypting, and storing the patients' voice over the semi-trusted server using voice-enabled IoT home devices, and voice-based search over mHealth data. Our scheme preserves both the richness and privacy of the voice data and achieves high efficiency in the voice search function.
	\item Third, we evaluate our schemes by performing privacy and accuracy analysis using real data and show our methods are successfully preserving the privacy of the data from the server, and accurately detect different tones, moods, and background sound from the collected voice data.
\end{itemize}

	The remainder of this paper is organized as follows; in section II we present the system model by introducing system components, design goals and, trust model. Section III covers the preliminaries of this work and followed by our proposed scheme at section IV, we bring the privacy and usability evaluation in section V. Then related works are presented in section VI and finally we conclude the paper in section VII.


\section{System Model}
We consider a scenario in which the patient has equipped her home with smart home devices and uses these devices to securely record and send her voice to her caregiver.

\subsection{System components}

\begin{figure}[!htbp]
	\includegraphics[width=.5\textwidth]{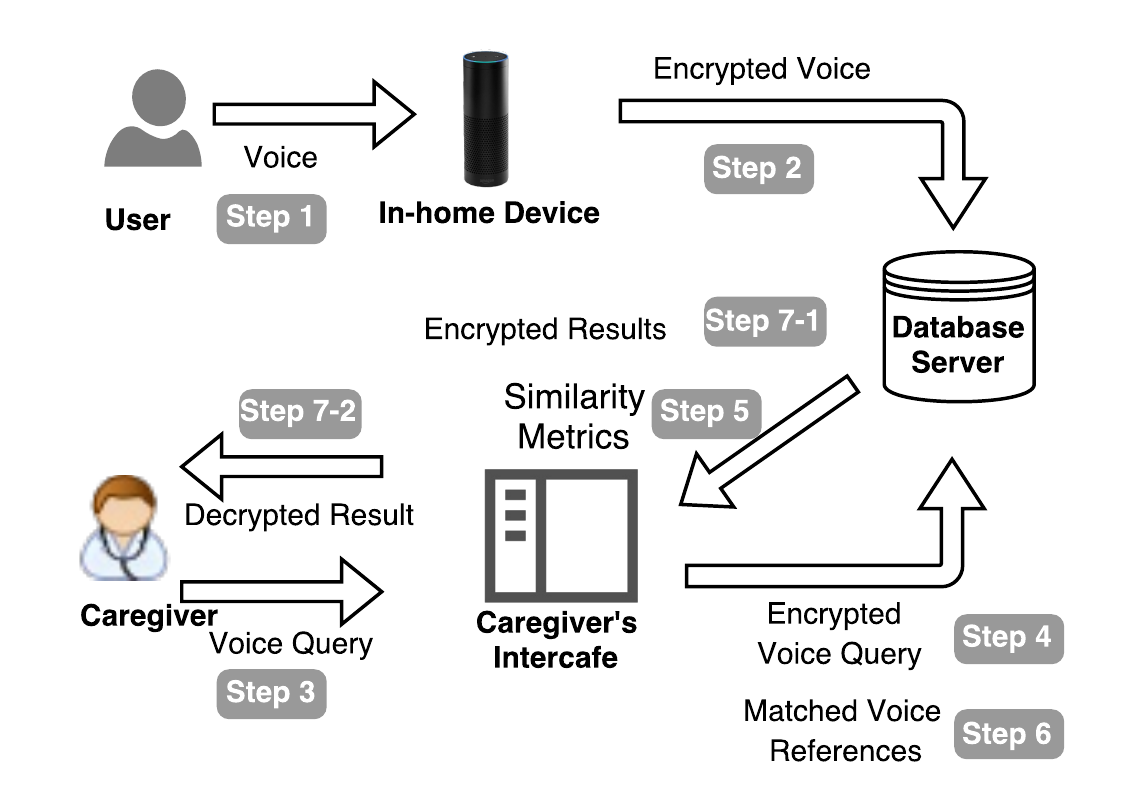}
	\vspace{-1cm}
	\caption{System Model}
	\vspace{-0.4cm}	
	\label{model}
\end{figure}

As depicted in Fig.~\ref{model}, our proposed system consists of five components:
\begin{itemize}
	\item User ($U$): the primary user, e.g. the patient who is using the system to communicate with her caregiver.
	\item Caregiver ($D$): the caregiver of the patient who uses the system to receive the recorded voice data and uses them to make queries and get similar voice data to study the mental and physical conditions of the patient.
	\item Device ($S$): the IoT home device which records, encrypts and uploads the user's voice data to a server.
	\item  Interface ($I$): the interface caregiver uses to get and decrypt the list of recorder voice data of the user, and also make voice search and queries on each of the samples.
	\item Database server ($DB$): the semi-trusted database server which is in charge of keeping the encrypted voice data and returning the query results after processing caregiver's encrypted query.
\end{itemize}

\subsection{Design goals}

\begin{itemize}
\item Voice richness preservation - Voice data contains more information than text, and these extra information are usually lost in current applications. In our system, voice data are not converted to text and the actual voices are being transmitted to the care giver, thus all the information contained in the data is being preserved and used.

\item Voice privacy preservation - Health related information are private and sensitive, and usually are required to be preserved from the untrusted servers. In our system, the server is considered semi-trusted, thus no information about the contents of the data being transmitted is shared with the server and only the user and her caregiver are aware of the contents of the information.

\item Search efficiency - The search and matching between the database contents and the given search voice are done by the server. This operation is done by comparing the features of the voice data. These features are only extracted once and stored along with the voice itself into the database, thus the server is able to efficiently match the given voices.

\end{itemize}
\subsection{Trust model}
\begin{itemize}
	\item Database server: Database server is semi-trusted in our system, i.e. the user information would not be shared with the server, but it is trusted to perform the voice matching tasks correctly and seamlessly.
	
	\item Home IoT device: The home device is considered trusted as it is collecting and processing user's sensitive and private voice information. It also is trusted to perform the encryption tasks on the recorded voices.
	
	\item Caregiver and her Interface: The caregiver of the patient is considered trusted and has the privileged to receive and use the sensitive voice information of the user. The interface device between the caregiver and the voice data returned from the server, which is used to receive and search the voice database is also considered trusted.
	
\end{itemize}

\section{Preliminaries}

\subsection{Homomorphic encryption}
Homomorphic encryption provides the addition and multiplication operations over ciphertexts, i.e. heavy operations can be performed by untrusted parties without knowing the shared secret. This method is widely used in data aggregation and computation on privacy-sensitive content~\cite{lu2012eppa}. A homomorphic encryption scheme can be described as follows:

A central authority runs a generator $\mathcal{G}$ which outputs $\langle p, q, R, R_q, R_p, \mathcal{X} \rangle$ as system public parameters: 

\begin{itemize}
\item $p\leq q$ are two primes s.t. $q \equiv 1 \mod 4$ and $p \gg l$;
\item Rings $R = \mathbb{Z}[x]/\langle x^2+1\rangle$, $R_q = R/qR = \mathbb{Z}_q[x]/\langle x^2+1\rangle$;
\item Message space $R_p = \mathbb{Z}_p[x] \langle x^2 + 1\rangle$;
\item A discrete Gaussian error distribution $\mathcal{X} = D_{Z^n,\sigma}$ with
standard deviation $\sigma$.
\end{itemize}

Suppose user $u_i$ has a public/private key pair $(pk_i, sk_i)$ such that $pk_i = \{a_i, b_i\}$, with $a_i = -(b_is + pe), b_i \in R_q$ and $s, e \leftarrow \mathcal{X}$, and $sk_i = s$. Let $b_{i,1}$ and $b_{i,2}$ be two messages encrypted by $u_i$.
\begin{itemize}
\item Encryption $E_{pk_i} (b_i,1)$: $c_{i,1} = (c_0, c_1)=(a_iu_t + pg_t +
b_{i,1}, b_iu_t + pf_t)$, where $u_t, f_t, g_t$ are samples from $\mathcal{X}$.
\item Decryption $D_{sk_i} (c_{i,1})$: If denoting $c_{i,1} = (c_0, ..., c_{\alpha_1} )$, then $b_{i,1} = (\sum_{k=0}^{\alpha_1}c_ks^k) \mod p$.\\
Consider the two ciphertexts $c_{i,1} = E(b_{i,1})=(c_0,..., c_{\alpha_1} )$ and $c_{i,2} = E(b_{i,2})=(c'_0,..., c'_{\alpha_1} )$.
\item Addition: Let $\alpha\ = \max(\alpha_1,\alpha_2)$. If $\alpha_1 \leq \alpha$, let $c_{\alpha_1+1}=... = c_{\alpha}= 0$; If$\alpha_2 \leq \alpha$, let $c'_{\alpha_2+1}=...=c'_\alpha=0$. Thus, we have $E(b_{i,1} + b_{i,2})=(c_0 \pm c'_0,..., c_\alpha \pm c'_\alpha)$.
\item Multiplication: Let $v$ be a symbolic variable and compute $(\sum_{k=0}^{\alpha_1}c_kv^k) . (\sum_{k=0}^{\alpha_2}c_kv^k) = \hat{c}_{\alpha_1+\alpha_2}v^{\alpha_1+\alpha_2}+...+\hat{c}_1v+\hat{c}_0$. Thus, we have $E(b_{i,1} \times b_{i,2}) = (\hat{c}_0,...,\hat{c}_{\alpha_1+\alpha_2})$.
\end{itemize}

\subsection{Voice feature extraction}
\begin{figure}[htbp]
\centering	\includegraphics[width=.45\textwidth]{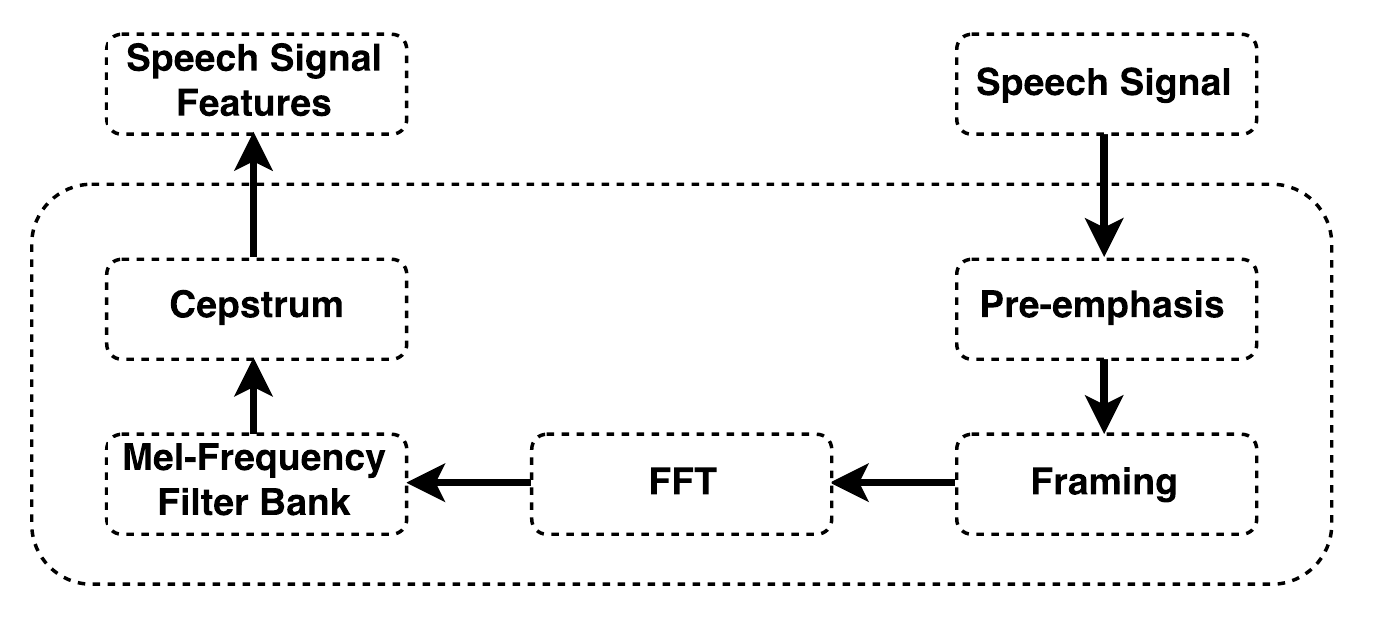}
	\vspace{-0.5cm}
	\caption{MFCC Overview}
	\vspace{-0.4cm}	
	\label{mfcc}
\end{figure}

Psychophysical studies~\cite{memon2009using} have shown that human perception of the sound frequency contents for speech signals does not follow a linear scale. Thus for each tone with an actual frequency, $f$, measured in Hz, a subjective pitch is measured on a scale called the ``Mel" scale.
\begin{center}
$f_{mel}  = \alpha\log_{10}(1+\frac{f}{700}) $
\end{center}
Where $f_{mel}$ is the subjective pitch in Mels corresponding to a frequency in Hz and $\alpha$ is a constant. This leads to the definition of ``Mel Frequency Cepstral Coefficient" (MFCC)\cite{han2006efficient}. Fig.~\ref{mfcc} depicts the block diagram for MFCC algorithm.

The voice signal is first pre-emphasized with a filter to spectrally flatten the signal. Then the pre-emphasized voice signal is separated into short segments called frames. There usually is a overlap between two adjacent frames ensure stationary between frames. Then a Fast Fourier Transform is applied to the frames and after that, the spectrum of each frame is filtered by a set of filters, and the power of each band is calculated. Finally we can calculate the Mel-Frequency cepstrum from the output power of the filter bank using this equation:
\begin{center}
	$c_n = \sum_{k=1}^{K} (\log S_k) \cos [n(k-0.5)\frac{\pi}{K}]$
\end{center}
\section{Proposed scheme}
In this section, we introduce our proposed data collection, encryption, and storage scheme as well as our proposed keyword search mechanism for query over encrypted voice data.

\subsection{Scheme overview}

The overview of our scheme is shown in Fig.~\ref{model}. Consider a scenario where the user has a voice-enabled IoT home device to communicate and transfer her voice data to her personal database accessible by her caregiver. The IoT home device collects and encrypts the user voice locally and stores the encrypted voice in the database. The database is always updated with the latest user health data records. Her caregiver, e.g. caregiver, using an interface containing the user's labeled voice samples which can send voice queries to the encrypted voice database, and obtain all the similar voices of that user using the same interface. This case cannot be addressed using the speech-recognition techniques because i) the tone of the patient's voice is also important for us, and ii)  there exists extra information on the background, which can reveal important information for the caregiver. There are six steps for the query process.

\textbf{Step 1:} Voice collection and encryption: User can easily and with the least effort record her voice using the voice enabled smart home device. This device collects and locally encrypts the user voice using a symmetric key encryption method~\cite{liang2013fully}.

\textbf{Step 2:} Encrypted voice upload: The in-home device then uploads the encrypted voice to the database server which is untrusted and cannot know the contents of the uploaded information, thus the encryption key is not shared with it.

\textbf{Step 3:} Query: User's caregiver, e.g. caregiver, queries the database using the labeling interface, which is an application which associates all the voice data in user's voice database to their suitable text labels defined by the caregiver. The encryption key is shared with this interface and caregiver can listen to the stored voices of the user, categorize and label them and then select the intended voices for the query.

\textbf{Step 4:} Query encryption: The selected keyword voice is then encrypted and sent to the server for the keyword matching. As mentioned before, since the server is untrusted, it does not have any information about the contents of the received query.

\textbf{Step 5:} Encrypted voice matching: Using our proposed encrypted voice matching mechanism, the server is able to calculate a similarity factor between the voice query and the suer voice data using the method introduced in subsection C. The server then returns these encrypted similarity metrics to the caregiver's interface.

\textbf{Step 6:} Similarity metric decryption and voice data request: Caregiver's interface then decrypts the received information from previous step and based on a pre-defined threshold values, detects a set of matching voices. References to these voices are then sent to the server for requesting the actual voice files.

\textbf{Step 7:} Results delivering: The voice data corresponding to the given references are returned from the server to caregiver's interfaces and then to the caregiver after decryption.

\subsection{Local voice encryption by home device}

Since the voice data to be transferred are personal and sensitive, and the server is considered untrusted, the voice files need to be encrypted locally before being transferred to the server. For this purpose, we first apply a Mel Frequency Cepstral Coefficient (MFCC) algorithm, as explained in the preliminaries section, on the voice files to extract the features needed for classifying the voices.

\begin{algorithm}
\small
\caption{MFCC Algorithm}\label{part}
\begin{algorithmic}[1]
\item \textbf{Input:} voice file $V$\\
\textbf{Output:} a 2-D array $F$, containing features of the voice based on the frequency\\
x=audioread($V$); \Comment {read the voice file} \\
bank=melbankm(rate($V$)); \Comment {create the Mel filter bank coefficients}\\
m = fft(x);  \Comment {Fast Fourier Transform}\\
dtm(i,:)=-2*m(i-2,:)-m(i-1,:)+m(i+1,:)+2*m(i+2,:);  \Comment {the first-order difference coefficient}\\
dtmm(i,:)=-2*dtm(i-2,:)-dtm(i-1,:)+dtm(i+1,:)+2*dtm(i+2,:);  \Comment {the second-order difference coefficient}\\
$F$ = [m dtm dtmm]; \Comment {merge all coefficients to get the features}
\textbf{return } $F$ \Comment{Features matrix}
\end{algorithmic}
\end{algorithm}
The feature matrix $F$, is a 36 by $l$, 2-d array of double values, where $l$ depends on the length of the input voice file and 36 is the number of filter banks we use. A sample output of this matrix is shown here:
\begin{equation} 
\begin{split}
v=
\begin{bmatrix}
6.960&	20.511&...&0.115\\
9.220&	21.134&...&-0.189\\
. & . & . & . \\
. & . & . & . \\
2.818&	8.825&...&-0.252\\
\end{bmatrix}
\end{split}
\end{equation}

The output of this algorithm is then encrypted using a homomorphic symmetric encryption method when the encryption key is shared only between the user and the caregiver. Also, the actual voice files themselves are encrypted, but not with the same encryption method because there is no operations performed on the voice data by the server. In this case we use AES encryption method for these voices. These encrypted voice samples are then associated with their encrypted features set and stored on the database on the cloud server together.

\subsection{Encrypted voice feature matching}
After the encrypted voice information from the user are stored on the database, now the caregiver would be able to query these information. To do so, all the voice samples are available to the caregiver, and since the patient shares the encryption key with the caregiver, she is able to decrypt and listen to the actual voice files recorded by the user one by one. Server will perform the following operations to match a received voice features $v'$ to all of the voice features $v$ in the database. Note that each voice sample consists of a $l$ by $36$ matrix from MFCC features as shown in equation~(2) where $l$ is depended on the length of each voice. For simplicity, the in-home device performs a column-based averaging operation and runs encryption on vector $v$ to obtian $E(\bar{v})$ as shown in equation~(\ref{ebarv}) where $E(x)$ stands for encrypted $x$.
\vspace{-0.15cm}
\begin{equation} \label{e(v)}
\begin{split}
v=
\begin{bmatrix}
v_{1,1} & v_{1,2} & ... &v_{1,36} \\
v_{2,1} & v_{2,2} & ... &v_{2,36} \\
. & . & . & . \\
. & . & . & . \\
v_{l,1} & v_{l,2} & ... &v_{l,36} \\
\end{bmatrix}
\end{split}
\end{equation}

\begin{equation} \label{ebarv}
E(\bar{v})= (E(\bar{v}_{1}) \ E(\bar{v}_{2}) \ ...\ E(\bar{v}_{36}) ) 
\end{equation}

Now, to calculate the similarity metric between the two voices, the server calculates $36$ element-by-element distances by calculating the squares of their subtraction and then gets an average on all the distance values to get a single value as the distance between two voices $v,v'$ as shown in equation~(\ref{distance}).

\begin{small}
\begin{equation} \label{distance}
E(dis_{v,v'})=
E((\bar{v}_{1}-\bar{v'}_{1})^2+(\bar{v}_{2}-\bar{v'}_{2})^2+...+(\bar{v}_{36}-\bar{v'}_{36})^2) 
\end{equation}
\end{small}
Note that all these operations on encrypted data are enabled because of the used homomorphic encryption. The caregiver's interface then decrypts this value and uses it to compare with pre-defined threshold values to decide if the two voice samples are similar enough. If they are, the caregiver further request the original voice sample using the references.
%
\section{Evaluation}
To evaluate the efficiency and privacy requirements of our proposed scheme, we have conducted experiments on real human and machine generated voice samples. We have used the online text-to-speech converters from ``https://acapela-box.com/" and also have recorded human voices using regular, i.e. not noise-canceling cellphone and computer microphones in general environments like homes, offices and university campuses with different background voices (the voice samples are publicly available for research purposes). The first part of the evaluation, privacy analysis, is to confirm the system satisfies the privacy requirements in order for the scheme to prevent the server to infer the contents of the voice samples. The second part confirms that the scheme is usable enough in terms of the server being able to match the stored voices and received voice queries, thus the scheme is actually working.

\subsection{Privacy analysis}
As mentioned in the proposed scheme, our scheme uses homomorphic encryption ($E$) to preserve the privacy of the user's sensitive voice features, i.e. $v$, and AES encryption ($Enc$) for voice data themselves, i.e. $voice$. That means only $Enc(voice)$ and $E(v)$ are stored on the server. The encryption keys are shared only between the user and the caregiver and server is not able to infer the information while processing the requests and queries. Homomorphic encryptions gives us the power of transferring the highly computational operations on the server without revealing the actual data to it, which is a big advantage in this case with sensitive information and highly computational operations. Moreover, while performing the matching operations on the encrypted voice features, all the intermediate and final results are still encrypted and there is no relationship between them inferable by the server.

\subsection{Efficiency evaluation}
The server must be able to accurately match the voice query with the stored voice data on the server. Specifically, we use two different thresholds learned from training data for two levels of similarity. Fist we will check if the voices are similar enough to be detected as same word with the same tone, or same background noise. This threshold value is called $T_m$. If this check fails, i.e. the scheme does not detect enough similarity between the tones or background noises, then the second threshold, which naturally is larger than the first one, is used to detect if the words are the same, but said in different moods. This threshold is called $T_w$. If the voices don't match even with this threshold, then they are considered different words. For instance, server should be able to match the voices containing the word ``happy", said in an excited tone with similar voices and tones, using the $T_m$, and it should be able to distinguish the word ``happy" said in a bored mood as the same word but a different mood, also using $T_m$. Formally, with the notation of $dis_{v,v'}$ for similarity between voice features $v$ and $v'$, and $\alpha.\beta$ for ``word.tone", following equations are always correct when $\alpha, \beta, \gamma$ are disparate:

\begin{small}
\begin{equation} \label{formal1}
dis_{\alpha.\beta,\alpha.\beta} \leq T_m
\end{equation}
\begin{equation} \label{formal2}
T_w \geq dis_{\alpha.\beta,\alpha.\gamma} \geq T_m\\
\end{equation}
\begin{equation} \label{formal3}
dis_{\alpha.\beta,\gamma.*} \geq T_w\\
\end{equation}
\end{small}
\vspace{-0.3cm}

Our results show that the user voice data in the database are distinguishable by the server with the accuracy of $86.66\%$ for same word with same mood, and $79.15\%$ for same word with two different moods. Also the similar background noises are distinguishable by the server with an accuracy of $76.66\%$. All the cases are shown in table \ref{usabletable}. We use the well-known ``Accuracy", ``Sensitivity" and ``Specificity" statistical metrics \footnote{Accuracy is defined as (True Positive + True Negative) / (True Positive + False Positive + True Negative + False Negative) , sensitivity is $TP / (TP + FN)$ and $Specificity = TN / (TN + FP)$.} to show our scheme's performance.
	\begin{table}
		\begin{tabular}{l l l l}
			\hline
			Similarity Level  & Accuracy & Sensitivity & Specificity\\
			\hline
			Same Mood & $86.66\%$ & 0.60 & 0.90 \\	
			Different Mood & $79.15\%$ & 0.44 & 0.92\\
			Different Background Noise& $76.66\%$ & 1.00 & 0.75\\
			\hline
		\end{tabular}
	\vspace{0.1cm}
		\caption{Same Word Efficiency Evaluation} \label{usabletable}
	\end{table}

\begin{figure}[htbp]
	\includegraphics[width=.5\textwidth]{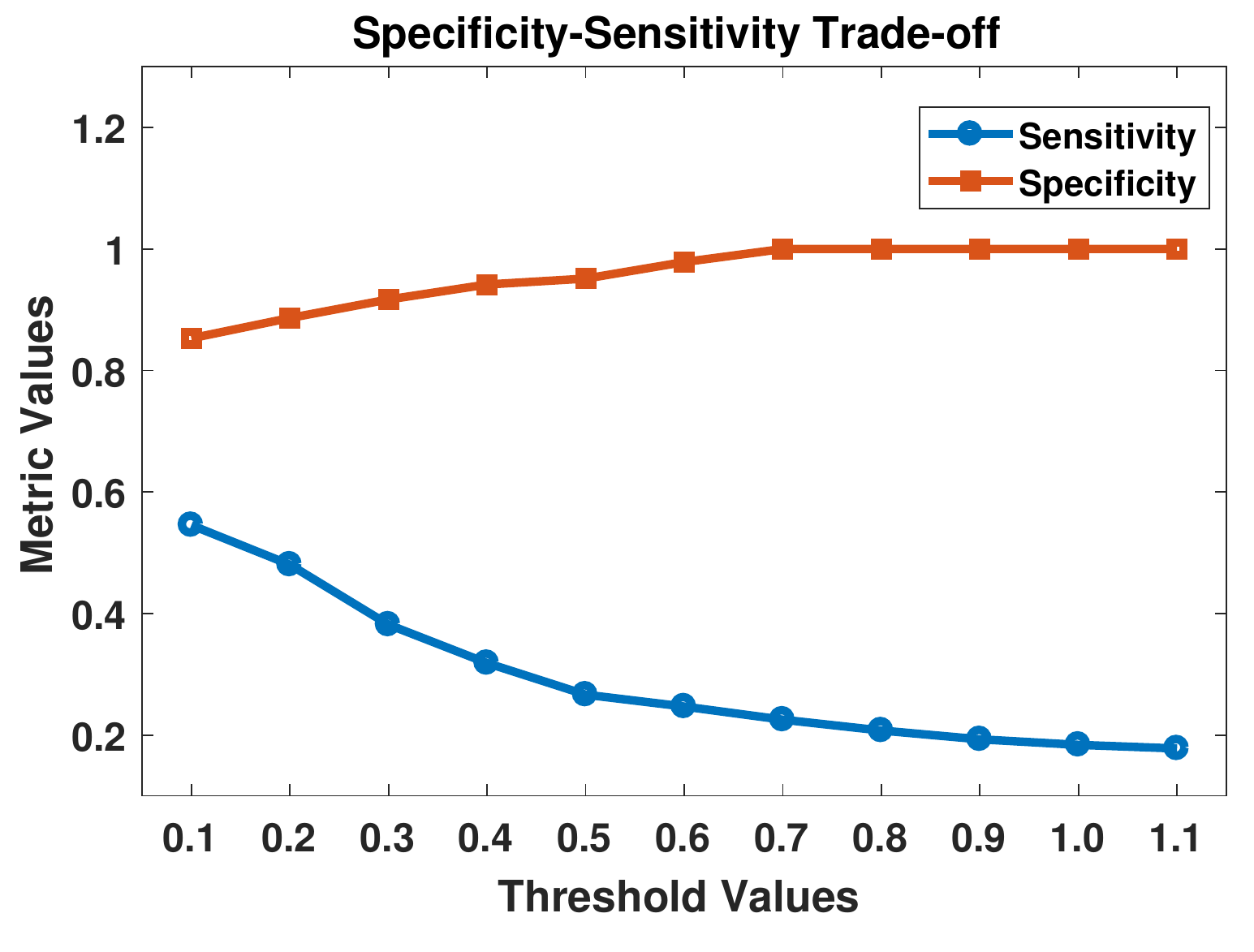}
	\vspace{-0.9cm}
	\caption{Specificity-Sensitivity Trade-off}
	\label{tradeoff}
\end{figure}

An important point to mention here is the relationship between specificity and sensitivity. Intuitively, for both thresholds, if we take the threshold too small, none of the voices would be detected as similar, i.e. number of false negatives will grow, but on the other hand number of false positives would decrease, which results in higher sensitivity and lower specificity. And vice versa for the case where we take the thresholds too large. This trade-off is depicted on the Fig.~\ref{tradeoff} for detection of the same word as an example.
\vspace{-0.5cm}
\vspace{-0.3cm}
\section{Related works}

IoT provides a perfect platform for smart ubiquitous healthcare~\cite{atzori2010internet} using body area sensors and IoT as the back-end for uploading the data. As an extension to body area sensors, the home monitoring systems for patient and specifically the elderly allows the caregivers to monitor the patients closely and continuously to avoid hospitalization costs~\cite{luo2010remote,nussbaum2006people}. However, the concept of voice-enabled home IoT devices, is a fairly new and topic and has not been addressed comprehensively.

With advances in speech recognition technologies, there are many voice enabled smart home devices introduced to help users to interact with devices via speech. However, there are several privacy and security concerns when dealing with microphone-enabled smart devices.~\cite{dhingra2013isolated} distinguishes active and passive listening and introduces three different categories; manually activated, speech activated, and always on and studies some privacy implications of these challenges and mentions the ability of these devices to passively listen to all the voices in the environment to detect their wake-up words.

Several prior works are done around data storage on untrusted cloud servers and query processing on encrypted data. Keyword search on encrypted data has been studied for many different utilities~\cite{fahrnberger2014sims,wang2010secure,li2016enabling,zhang2011efficient,duan2015privacy,naveed2014dynamic}. While standard methods of secrecy hide the content of the message, covert communication in wireless environments~\cite{soltani2014covert, soltani2018covert} and computer networks~\cite{soltani2015covert, soltani2016covert} hides the existence of the communication. More Specifically,~\cite{shafagh2015talos} proposed an encryption method for IoT data storage and query processing on untrusted cloud database servers which applies the encryption at the origin of the data, but the server is still able to process the queries, however these studies are more concentrated on text as the type of stored data and voice data types are not widely considered for this purpose.

Mel Frequency Cepstral Coefficient (MFCC) has been widely used for extracting the features of spoken human voice mainly for speech recognition purposes~\cite{dhingra2013isolated,muda2010voice,bala2010voice,goh2009robust,martinez2012speaker} but this technique can also be applied for matching the features of the voice to associated voices directly without conversion to text.

Speech recognition and voice-to-text services are studied and implemented in a wide variety of applications as a promising technique to minimize the storage requirements and facilitate the data processing and information extraction. However, the requirement of a powerful processing capability has always been a drawback. Recently, local speech recognition engines are introduced~\cite{dhingra2013isolated,mcgraw2016personalized}, but the capability of such systems are low and learning and customization of the system based on the individuals' speech is very limited~\cite{mcgraw2016personalized}. On the other hand, using cloud computing for this purpose will require disclosure of the information to the server, which may be untrusted. Our proposed scheme addresses these issues by replacing speech recognition by a novel encrypted voice matching technique.
\section{Conclusion}
In this paper, we proposed an efficient and privacy-preserving voice-based search scheme. We studied the importance and advantages of the extra information in the voice over text for healthcare applications. We employed the voice feature extraction and matching algorithms to achieve the matching efficiency, and we employed the homomorphic encryption technique to achieve the voice privacy.  Through evaluation in real experiment we showed that our scheme is able to detect the tone and background voices in the patient's recorded voice data and categorize the voices based on them with an average accuracy of 80.8\%. For our future work, we will include more characteristics of voice data in our scheme design such that the caregiver can make voice-based search with these characteristics. We will also explore other techniques for voice feature extraction to achieve higher accuracy.
\section{Acknowledgment}
This research program is supported by Joseph P. Healey Research Grant from UMass Boston, National Science Foundation award number 1618893, and National Science Foundation of Fujian province China (Grant No. 2016J01325).

\bibliographystyle{IEEEtran}
\bibliography{ref}

\begin{thebibliography}{10}
\providecommand{\url}[1]{#1}
\csname url@samestyle\endcsname
\providecommand{\newblock}{\relax}
\providecommand{\bibinfo}[2]{#2}
\providecommand{\BIBentrySTDinterwordspacing}{\spaceskip=0pt\relax}
\providecommand{\BIBentryALTinterwordstretchfactor}{4}
\providecommand{\BIBentryALTinterwordspacing}{\spaceskip=\fontdimen2\font plus
\BIBentryALTinterwordstretchfactor\fontdimen3\font minus
  \fontdimen4\font\relax}
\providecommand{\BIBforeignlanguage}[2]{{%
\expandafter\ifx\csname l@#1\endcsname\relax
\typeout{** WARNING: IEEEtran.bst: No hyphenation pattern has been}%
\typeout{** loaded for the language `#1'. Using the pattern for}%
\typeout{** the default language instead.}%
\else
\language=\csname l@#1\endcsname
\fi
#2}}
\providecommand{\BIBdecl}{\relax}
\BIBdecl

\bibitem{holopainen2007use}
A.~Holopainen, F.~Galbiati, and K.~Voutilainen, ``Use of smart phone
  technologies to offer easy-to-use and cost-effective telemedicine services,''
  in \emph{Digital Society, 2007. ICDS'07. First International Conference on
  the}.\hskip 1em plus 0.5em minus 0.4em\relax IEEE, 2007, pp. 4--4.

\bibitem{ceer2006pervasive}
D.~Ceer, ``Pervasive medical devices: less invasive, more productive,''
  \emph{IEEE Pervasive Computing}, vol.~5, no.~2, pp. 85--87, 2006.

\bibitem{schober2015precision}
M.~F. Schober, F.~G. Conrad, C.~Antoun, P.~Ehlen, S.~Fail, A.~L. Hupp,
  M.~Johnston, L.~Vickers, H.~Y. Yan, and C.~Zhang, ``Precision and disclosure
  in text and voice interviews on smartphones,'' \emph{PloS one}, vol.~10,
  no.~6, p. e0128337, 2015.

\bibitem{forsberg2003speech}
M.~Forsberg, ``Why is speech recognition difficult,'' \emph{Chalmers University
  of Technology}, 2003.

\bibitem{online}
J.~Halamka, ``{Early Experiences with Ambient Listening Devices (Alexa and
  Google Home)},''
  \url{http://geekdoctor.blogspot.com/2017/03/early-experiences-with-ambient.html},
  2017, [Online; accessed 20-April-2017].

\bibitem{mcgraw2016personalized}
I.~McGraw, R.~Prabhavalkar, R.~Alvarez, M.~G. Arenas, K.~Rao, D.~Rybach,
  O.~Alsharif, H.~Sak, A.~Gruenstein, F.~Beaufays \emph{et~al.}, ``Personalized
  speech recognition on mobile devices,'' in \emph{Acoustics, Speech and Signal
  Processing (ICASSP), 2016 IEEE International Conference on}.\hskip 1em plus
  0.5em minus 0.4em\relax IEEE, 2016, pp. 5955--5959.

\bibitem{boneh2004public}
D.~Boneh, G.~Di~Crescenzo, R.~Ostrovsky, and G.~Persiano, ``Public key
  encryption with keyword search,'' in \emph{International Conference on the
  Theory and Applications of Cryptographic Techniques}.\hskip 1em plus 0.5em
  minus 0.4em\relax Springer, 2004, pp. 506--522.

\bibitem{lu2012eppa}
R.~Lu, X.~Liang, X.~Li, X.~Lin, and X.~Shen, ``Eppa: An efficient and
  privacy-preserving aggregation scheme for secure smart grid communications,''
  \emph{IEEE Transactions on Parallel and Distributed Systems}, vol.~23, no.~9,
  pp. 1621--1631, 2012.

\bibitem{memon2009using}
S.~Memon, M.~Lech, and L.~He, ``Using information theoretic vector quantization
  for inverted mfcc based speaker verification,'' in \emph{Computer, Control
  and Communication, 2009. IC4 2009. 2nd International Conference on}.\hskip
  1em plus 0.5em minus 0.4em\relax IEEE, 2009, pp. 1--5.

\bibitem{han2006efficient}
W.~Han, C.-F. Chan, C.-S. Choy, and K.-P. Pun, ``An efficient mfcc extraction
  method in speech recognition,'' in \emph{Circuits and Systems, 2006. ISCAS
  2006. Proceedings. 2006 IEEE International Symposium on}.\hskip 1em plus
  0.5em minus 0.4em\relax IEEE, 2006, pp. 4--pp.

\bibitem{liang2013fully}
X.~Liang, X.~Li, K.~Zhang, R.~Lu, X.~Lin, and X.~S. Shen, ``Fully anonymous
  profile matching in mobile social networks,'' \emph{IEEE Journal on Selected
  Areas in Communications}, vol.~31, no.~9, pp. 641--655, 2013.

\bibitem{atzori2010internet}
L.~Atzori, A.~Iera, and G.~Morabito, ``The internet of things: A survey,''
  \emph{Computer networks}, vol.~54, no.~15, pp. 2787--2805, 2010.

\bibitem{luo2010remote}
H.~Luo, S.~Ci, D.~Wu, N.~Stergiou, and K.-C. Siu, ``A remote markerless human
  gait tracking for e-healthcare based on content-aware wireless multimedia
  communications,'' \emph{IEEE Wireless Communications}, vol.~17, no.~1, 2010.

\bibitem{nussbaum2006people}
G.~Nussbaum, ``People with disabilities: assistive homes and environments,''
  \emph{Computers Helping People with Special Needs}, pp. 457--460, 2006.

\bibitem{dhingra2013isolated}
S.~D. Dhingra, G.~Nijhawan, and P.~Pandit, ``Isolated speech recognition using
  mfcc and dtw,'' \emph{International Journal of Advanced Research in
  Electrical, Electronics and Instrumentation Engineering}, vol.~2, no.~8, pp.
  4085--4092, 2013.

\bibitem{fahrnberger2014sims}
G.~Fahrnberger, ``Sims: A comprehensive approach for a secure instant messaging
  sifter,'' in \emph{Trust, Security and Privacy in Computing and
  Communications (TrustCom), 2014 IEEE 13th International Conference on}.\hskip
  1em plus 0.5em minus 0.4em\relax IEEE, 2014, pp. 164--173.

\bibitem{wang2010secure}
C.~Wang, N.~Cao, J.~Li, K.~Ren, and W.~Lou, ``Secure ranked keyword search over
  encrypted cloud data,'' in \emph{Distributed Computing Systems (ICDCS), 2010
  IEEE 30th International Conference on}.\hskip 1em plus 0.5em minus
  0.4em\relax IEEE, 2010, pp. 253--262.

\bibitem{li2016enabling}
H.~Li, Y.~Yang, T.~H. Luan, X.~Liang, L.~Zhou, and X.~S. Shen, ``Enabling
  fine-grained multi-keyword search supporting classified sub-dictionaries over
  encrypted cloud data,'' \emph{IEEE Transactions on Dependable and Secure
  Computing}, vol.~13, no.~3, pp. 312--325, 2016.

\bibitem{zhang2011efficient}
B.~Zhang and F.~Zhang, ``An efficient public key encryption with
  conjunctive-subset keywords search,'' \emph{Journal of Network and Computer
  Applications}, vol.~34, no.~1, pp. 262--267, 2011.

\bibitem{duan2015privacy}
X.~Duan, J.~He, P.~Cheng, Y.~Mo, and J.~Chen, ``Privacy preserving maximum
  consensus,'' in \emph{Decision and Control (CDC), 2015 IEEE 54th Annual
  Conference on}.\hskip 1em plus 0.5em minus 0.4em\relax IEEE, 2015, pp.
  4517--4522.

\bibitem{naveed2014dynamic}
M.~Naveed, M.~Prabhakaran, and C.~A. Gunter, ``Dynamic searchable encryption
  via blind storage,'' in \emph{Security and Privacy (SP), 2014 IEEE Symposium
  on}.\hskip 1em plus 0.5em minus 0.4em\relax IEEE, 2014, pp. 639--654.

\bibitem{soltani2014covert}
R.~Soltani, B.~Bash, D.~Goeckel, S.~Guha, and D.~Towsley, ``Covert single-hop
  communication in a wireless network with distributed artificial noise
  generation,'' in \emph{Communication, Control, and Computing (Allerton), 2014
  52nd Annual Allerton Conference on}.\hskip 1em plus 0.5em minus 0.4em\relax
  IEEE, 2014, pp. 1078--1085.

\bibitem{soltani2018covert}
R.~Soltani, D.~Goeckel, D.~Towsley, B.~Bash, and S.~Guha, ``Covert wireless
  communication with artificial noise generation,'' \emph{IEEE Transactions on
  Wireless Communications}, pp. 1--1, 2018.

\bibitem{soltani2015covert}
R.~Soltani, D.~Goeckel, D.~Towsley, and A.~Houmansadr, ``Covert communications
  on poisson packet channels,'' in \emph{Communication, Control, and Computing
  (Allerton), 2015 53rd Annual Allerton Conference on}.\hskip 1em plus 0.5em
  minus 0.4em\relax IEEE, 2015, pp. 1046--1052.

\bibitem{soltani2016covert}
------, ``Covert communications on renewal packet channels,'' in
  \emph{Communication, Control, and Computing (Allerton), 2016 54th Annual
  Allerton Conference on}.\hskip 1em plus 0.5em minus 0.4em\relax IEEE, 2016,
  pp. 548--555.

\bibitem{shafagh2015talos}
H.~Shafagh, A.~Hithnawi, A.~Dr{\"o}scher, S.~Duquennoy, and W.~Hu, ``Talos:
  Encrypted query processing for the internet of things,'' in \emph{Proceedings
  of the 13th ACM Conference on Embedded Networked Sensor Systems}.\hskip 1em
  plus 0.5em minus 0.4em\relax ACM, 2015, pp. 197--210.

\bibitem{muda2010voice}
L.~Muda, M.~Begam, and I.~Elamvazuthi, ``Voice recognition algorithms using mel
  frequency cepstral coefficient (mfcc) and dynamic time warping (dtw)
  techniques,'' \emph{arXiv preprint arXiv:1003.4083}, 2010.

\bibitem{bala2010voice}
A.~Bala, A.~Kumar, and N.~Birla, ``Voice command recognition system based on
  mfcc and dtw,'' \emph{International Journal of Engineering Science and
  Technology}, vol.~2, no.~12, pp. 7335--7342, 2010.

\bibitem{goh2009robust}
C.~Goh and K.~Leon, ``Robust computer voice recognition using improved mfcc
  algorithm,'' in \emph{New Trends in Information and Service Science, 2009.
  NISS'09. International Conference on}.\hskip 1em plus 0.5em minus 0.4em\relax
  IEEE, 2009, pp. 835--840.

\bibitem{martinez2012speaker}
J.~Martinez, H.~Perez, E.~Escamilla, and M.~M. Suzuki, ``Speaker recognition
  using mel frequency cepstral coefficients (mfcc) and vector quantization (vq)
  techniques,'' in \emph{Electrical Communications and Computers (CONIELECOMP),
  2012 22nd International Conference on}.\hskip 1em plus 0.5em minus
  0.4em\relax IEEE, 2012, pp. 248--251.

\end{thebibliography}
%

\end{document}